\documentstyle[12pt]{article}

\def\be{\begin{equation}}
\def\ee{\end{equation}}
\def\ba{\begin{array}{c}}
\def\ea{\end{array}}

\def\ben{\[}
\def\een{\]}
\begin{document}

\titlepage
%\vspace*{2cm}

 \begin{center}{\Large \bf
Partial sums and optimal shifts in

shifted large$-\ell$ perturbation expansions

 for quasi-exact potentials
  }\end{center}

%\vspace{5mm}

 \begin{center}
Miloslav Znojil

%\vspace{5mm}

 Nuclear Physics Institute AS {C}R,
250 68 \v{R}e\v{z}, CZECH REPUBLIC

{e-mail: znojil@ujf.cas.cz }

\vspace{5mm}

\end{center}

\vspace{3mm}

\section*{Abstract}

For the $N-$plets of bound states in a quasi-exactly solvable (QES)
toy model (sextic oscillator), the spectrum is known to be given as
eigenvalues of an $N$ by $N$ matrix. Its determination becomes purely
numerical for all the larger $N>N_0=9$.  We propose a new
perturbative alternative to this construction. It is based on the
fact that at any $N$, the problem turns solvable in the limit of very
large angular momenta $\ell\to\infty$.  For all the finite $\ell$ we
are then able to define the QES spectrum by convergent perturbation
series. These series admit a very specific rational resummation,
having an analytic or branched continued-fraction form at the
smallest $N=4$ and $5$ or $N=6$ and $7$, respectively. It is
remarkable that among all the asymptotically equivalent small
expansion parameters $\mu \sim 1/(\ell + \beta)$, one must choose an
optimal one, with unique shift $\beta=\beta(N)$.

 \vspace{3mm}

\noindent
 AMS (MSC 2000): 81Q15

 \vspace{3mm}

\noindent
 Keywords:

 sextic oscillators;
 exact solvability;
 convergent perturbation series;
   generalized continued fractions

%\vspace{9mm}

% \begin{center}
%{\small \today, shif.tex file}
%\end{center}

\newpage

\section{Introduction \label{prvni}}

The $D-$dimensional and central Schr\"{o}dinger bound-state problem
 \[
 \left [ -\triangle +
 V(|\vec{r}|)
 \right ]
  \Psi(\vec{r})={\cal E}\,
  \Psi(\vec{r})
 \label{CSE}
 \]
finds its applications as an approximate model in nuclear physics
\cite{Sotona} as well as in quantum chemistry \cite{Dunn} and atomic
physics \cite{Goodson}. Last but not least, it is
encountered as a popular exercise in textbooks on quantum
mechanics~\cite{Constantinescu}. In all these cases, the main {\em
mathematical} merit of the model (\ref{CSE}) lies in its
separability, i.e., reducibility to the set
of the {\em ordinary} (so called radial) differential equations
 \begin{equation}
\left[-\,\frac{d^2}{dr^2} + \frac{\ell(\ell+1)}{r^2}+
 V(r)\right]\,
\psi(r) = {\cal E}\, \psi(r)\,,\ \ \ \ \ \ \ell=(D-3)/2+m\,,\ \ \ \ \
m = 0, 1, \ldots\,.
 \label{SE}
 \end{equation}
They are all easily solvable by any standard numerical
technique~\cite{ctverne}.

Whenever we need a non-numerical solution, we must restrict the class
of our potentials $V(|\vec{r}|)$ to a subset which is exactly
solvable (ES, say, harmonic-oscillator) or at least quasi-exactly
solvable (QES, see, e.g., the monograph \cite{Ushveridze} for a
thorough review).  Of course, the real appeal of any ES or QES
potential derives from an agreement of its shape or spectrum with
some ``more realistic" numerical model $V^{[phys]}(|\vec{r}|)$. In
this sense, many practical applications of the ES subset are marred
by its comparatively small size and, hence, by a very narrow
variability of the available shapes of $V^{[ES]}(|\vec{r}|)$ (cf.,
e.g., ref. \cite{CKS} for their representative list).  In parallel,
the wealth of the forms of the multi-parametric wells
$V^{[QES]}(|\vec{r}|)$ may happen to be more than compensated by the
much more complicated tractability of their
perturbations~\cite{Czech}.  Moreover, even the very zero-order QES
constructions may become quite complicated, whenever the size $N$ of
the required QES multiplet leaves the domain of $N \leq N_0$ where
the practical determination of the spectrum remains non-numerical
\cite{Leach}. Although the latter remark sounds like a paradox, one
must keep in mind that even in the simplest QES models, the explicit
construction of the larger QES multiplets with $N > N_0$ requires the
more or less purely numerical diagonalization of asymmetric $N$ by
$N$ matrices (see below). In purely pragmatic terms, many closed QES
constructions \cite{quartic,Gerdt} are in fact more difficult than an
immediate numerical solution of the differential Schr\"{o}dinger
eq.~(\ref{SE}) itself.

In what follows we shall argue that it is possible to avoid the
latter brute-force methods even in the QES domain where $N > N_0$. We
shall recommend the use of the so called large$-\ell$ (or large$-D$
\cite{dis}) expansion philosophy \cite{Chatterjee} in its new
implementation which takes into account several specific features of
QES models. We shall reduce all the inessential technicalities to a
necessary minimum and restrict our attention just to one of the most
elementary illustrative QES examples
 \be
 V^{[QES]}(|\vec{r}|) =
  a\,|\vec{r}|^2
 + c\,|\vec{r}|^6, \ \ \ \ \ \  c=\gamma^2 > 0
  \label{toy}
 \ee
which is known to possess an elementary $N-$plet of QES solutions at
any $N$ \cite{Singh}.  On the background of this example we shall
start our considerations in section \ref{druha} by showing that due
to the absence of the quartic term in eq. (\ref{toy}), the QES
solvability conditions are transparent and instructive (cf.
subsection \ref{prvnidruha}). In subsection \ref{druhadruha} we show
how these conditions simplify even the standard large$-\ell$
perturbation expansions of the spectrum where one uses just an
asymptotic, $\ell \gg 1$ simplification of the {\em differential}
Schr\"{o}dinger equation (\ref{SE}).  In section \ref{tretidruha} we
add that in the more specific, QES-related setting, the {\em
algebraic} construction of the solutions might be preferrable, at
least due to its amazingly transparent form.

The latter expectations are more than confirmed in the subsequent
section~\ref{treti}.  After a step-by-step analysis of the QES
secular equations for the energies in dependence on the growing
dimension $N$, we are able to reveal their general polynomial
structure and perturbation form. The first subsection
\ref{prvnitreti} shows that the models with $N\leq 3$ are all fully
solvable via an elementary re-scaling of the matrix secular equation.
Formally, our presently conjectured ``optimally shifted" large$-\ell$
series remain trivial in this case. The zero-order result is exact
and its secular polynomial does not contain any perturbation.  All
the higher-order corrections simply vanish.  Still, the structure of
the $N=3$ secular equation is already rich enough to illustrate, why
the ``optimal" value of the shift of $\ell \gg 1$ is so exceptional,
and how a return to its ``non-optimal" values would re-introduce
non-vanishing perturbation corrections.

In subsection \ref{druhatreti}, some of these observations are
re-confirmed at $N=4$. Elementary analysis offers the explanation why
our ``optimally shifted" large$-\ell$ power series are convergent.
The determination of the explicit value of the radius of convergence
$\mu_{max}$ reveals that the circle of convergence covers the {\em
whole} domain of physical interest, i.e., formally, all the positive
spatial dimensions $D>0$.

Subsection \ref{tretitreti} uses the slightly modified $N=5$ example
and shows that a Pad\'{e}-like re-summation of our power series
exists.  The ``re-summed" energies $E$ (as well as their squares
$\Omega=E^2$) acquire an analytic continued-fraction form.

In subsection \ref{ctvrtatreti}, the ``first nontrivial" example is
shown to emerge at $N=6$. We sample there the energies in their
``generalized" or ``branched" continued-fraction form as well as in
their more standard power-series representation. A few questions
concerning their convergence are clarified.

In subsection \ref{patatreti}, an extrapolation of all the preceding
constructions to any size $N$ of the QES multiplets is described and
discussed.  For quadruplets of $N = 4K, 4K+1, 4k+2$ and $4K+3$, our
main result is formulated as an iterative $K-$term formula for the
systematic construction of the energies in a generalized
continued-fraction form.  Simultaneously, the Taylor-expansion
technique is shown to generate the standard power series with the
``optimally shifted" large$-\ell$ perturbation structure.  The last
illustrations are added showing the smoothness of transition from the
``last solvable" $N \leq N_0=9$ to ``the first unsolvable" $N=10$.

In section \ref{ctvrta} a thorough re-interpretation of our optimally
shifted large$-\ell$ perturbation series is given in the more
standard language of the textbook Rayleigh-Schr\"{o}dinger
perturbation theory. Once more we summarize our approach in the last
section \ref{sesta}, as a source of new types of perturbation
expansions tailored for the computation of the QES spectra at any
multiplet size~$N$.

\section{Conditions of quasi-exact solvability
 \label{druha}}

\subsection{Toy model: sextic oscillator
\label{prvnidruha}}

Sextic oscillator (\ref{toy}) may
support elementary bound states
 \begin{equation}
\psi(r) = \sum_{n=0}^{N-1}\, h_n \,r^{2n+\ell+1}\,{\rm exp}\left( -
\frac{1}{4}{\gamma\,r^4 }\right)\,
 \label{ana}
 \end{equation}
provided only that we choose one of the couplings in
consistent manner~\cite{Singh},
 \be
  a=a(N)=
 -\gamma\,(4N+2\ell+1)\,.
 \label{cond}
 \ee
The latter condition increases a phenomenological appeal of our
example by assigning to  potential (\ref{toy}) the manifestly
non-perturbative double-well shape. Many computational difficulties
of ``realistic" calculations may be mimicked, even within such a
framework, whenever one chooses a larger degree $N$ in wave function
(\ref{ana}).  Integer $N$ measures the ``dimension" of our problem
since, under the constraint (\ref{cond}) and after the insertion of
(\ref{ana}), our differential Schr\"{o}dinger equation (\ref{SE}) is
transformed into equivalent matrix problem
 \begin{equation}
 \left(
  \begin{array}{ccccc}
 0 & C_0&  & &  \\
A_1&0 & C_1&    & \\ &\ddots&\ddots&\ddots&\\ &&A_{N-2}&0&C_{N-2}\\
&&&A_{N-1}&0
 \end{array}
 \right )\left( \begin{array}{c}
 {h}_0\\
 {h}_1\\
\vdots \\
 {h}_{N-2}\\
 {h}_{N-1}
\end{array} \right )={\cal E}\,\left( \begin{array}{c}
 {h}_0\\
 {h}_1\\
\vdots \\
 {h}_{N-2}\\
 {h}_{N-1}
\end{array} \right )\,.
 \label{tridSE}
 \end{equation}
The matrix of this system is asymmetric,
 \begin{equation}
 A_n = -4\,\gamma\,(N-n)  , \ \ \ \ \ \ \
 C_n = -2(n+1)\,(2n+2\ell+3),
  \ \ \ \ n = 0, 1, \ldots \
  \label{elem2}
   \end{equation}
so that up to the first few lowest dimensions $N$, equation
(\ref{tridSE}) need not be easy to solve at all. The advantages
gained by the polynomiality of $\psi$ may be quickly lost with the
growth of their degree $N$.  The merits of the ``exceptional" QES
levels seem to disappear in the domain of the larger $N \gg 1$.
However, these states may ``remember" and share some features of the
completely solvable harmonic oscillator. In this sense, our
oversimplified toy example may serve as a guide towards a future
better understanding of the partially solvable Schr\"{o}dinger
equations with more free parameters~\cite{Gerdt}.

\subsection{Large$-\ell$ domain in coordinate representation
\label{druhadruha}}

For our particular QES model of section \ref{prvnidruha}, the radial
Schr\"{o}dinger eq. (\ref{SE}),
 \begin{equation}
\left[-\,\frac{d^2}{dr^2} + \frac{\ell(\ell+1)}{r^2}
-\gamma\,(4N+2\ell+1)\,{r}^2
 + \gamma^2\,{r}^6\right]\,
\psi(r) = {\cal E}\, \psi(r)\,,
 \label{SExt}
 \end{equation}
may be treated by the so called $1/{\cal N}$ (i.e., in the
present notation, $1/\ell$) expansions along the lines recommended
in the  recent review \cite{Bjerrum}. This approach visualizes the
Hamiltonian in eq. (\ref{SExt}) as composed of an
``effective" kinetic energy in one dimension and of an ``effective"
one-dimensional potential well $V_{eff}(r)$ with a steep and
repulsive peak near the origin and also with a
confining growth at large $r$. Its minimum is unique and may be
localized at the point
 \be
 r = R(\ell)=
 \left (
 {\frac {4\,N+2\,{\ell}+1 +\triangle(\ell)} {{ 6\,\gamma}}}
 \right )^{1/4}
 ,
 \ee
 \ben
 \triangle(\ell) =
\sqrt {16\,{N}^{2}+16\,N{ \ell}+8\,N+16\,{{ \ell}}^{2}+16\,{ \ell}+1}
 . \een
At an illustrative strength $\gamma=1$ and for the simplest $N=0$ we
have
 \ben
 V_{eff}(r) =
 \left (
 \frac{\ell^2}{r^2} - 2\ell\,r^2 + r^6
 \right ) \,[ 1 + {\cal O}(1/\ell) ]
 = 16\,\ell\,[r-R(\ell)]^2+
 16\,\ell^{3/4}\,[r-R(\ell)]^3 + \ldots\,.
 \een
As long as the position of the minimum grows with $\ell$ (or $D$) to
its asymptotically leading-order value $R(\ell) =
(\ell/\gamma)^{1/4}(1 + {\cal O}(1/\ell)$ (note that we switched to
$\gamma=1$ temporarily), we may abbreviate
$r-R(\ell)=\xi/(2\,\ell^{1/4})$ and replace our Schr\"{o}dinger
equation (\ref{SExt}) by its $\ell \gg 1$ asymptotically equivalent
anharmonic-oscillator form
 \begin{equation}
 \left[-\,\frac{d^2}{d \xi^2} +
 \xi^2+ \frac{1}{2\sqrt{\ell}}\,\xi^3+ \frac{5}{16\,\ell}\,\xi^4 + {\cal
 O}
 \left ( \frac{\xi^5}{\ell^{3/2}} \right )
 \,\right]\, \phi(\xi) = \frac{{\cal E}}{4\,\sqrt{\ell}}\, \phi(\xi)\,
 \label{SExtas}
 \end{equation}
In terms of perturbation theory, the leading-order
energies coincide with a re-scaled equidistant spectrum of harmonic
oscillator,
 \be
 {\cal E} = 4\,\sqrt{\ell}\,(2n+1)\,[1 + {\cal
O}(1/\sqrt{\ell})], \ \ \ \ \ \ n = 0, 1, \ldots\,
 \ee
All the higher-order corrections may be computed in
systematic manner, leading to an asymptotic, divergent
\cite{Bjerrum} power series in $\lambda ={\cal
O}(1/\sqrt{\ell})$. In our present paper, we are going to
replace them by convergent series in the powers of $\lambda^4$.

\subsection{Large$-\ell$ domain in matrix representation
\label{tretidruha}}

Partially, our present paper has been motivated by all the realistic
models (\ref{SE}) where the dimension $D$ is large enough [like,
e.g., in ref. \cite{Sotona} where $D={\cal O}(10^3)$]. In the
preceding subsection we also saw that there might exist correlations
between the changes of $D$ (or $\ell$) and of the conditions of
(quasi-) exact solvability. Let us now pay more attention to this
phenomenon which appeared as a source of inspiration in several
studies of QES problems \cite{Dubna} where it has been noticed that,
at a fixed {\em matrix} dimension $N$, the solutions are getting
simpler whenever the {\em spatial} dimension $D$ grows to infinity.

In an introductory step we imagine that the QES  constraint
(\ref{cond}) merely replaces the definition (\ref{elem2}) of the
lower diagonal in eqs. (\ref{tridSE}) by the shorter formula
$A_n=4\gamma\,(n-N)$. An uncomfortable asymmetry of our linear
algebraic problem $Q({\cal E})\vec{h}=0$ (\ref{tridSE}) is still
there but it may be weakened by the shift of the large integer
$2\ell=D+2m-3$ (in the $m-$th partial wave) to another (and also
large) auxiliary integer $G = D+2m+N-2$. This shift represents
$Q({\cal E})$ as a slightly more balanced matrix dominated by its
upper diagonal,
 \ben
 \left(  \begin{array}{ccccc}
\ {\cal E} & 2(G+2-N) & & &  \\
 4(N-1)\gamma & {\cal E}\ \ \ \ \ \  & 4(G+4-N) &  &  \\
 &&&&\\
 &\ddots&\ddots\ \ \ \ \ \ \  &\ddots&\\
 &&&&\\
&&\ \  8\gamma & {\cal E}\ & 2 ({N-1})(G-2+N) \\ &&& 4\gamma & {\cal
E} \ \ \ \ \ \ \ \
\end{array} \right).
 \label{tr}
 \een
We re-scale it by its pre-multiplication by the diagonal matrix
$\hat{\rho}$ with elements $\rho^{j}$ where $\rho
=\sqrt{G/(2\gamma)}$. The parallel renormalization of the Taylor
coefficients $h_j \to {p}_j=[G/(2\gamma)]^{j/2}h_j$ enables us to
re-scale the energy,
  \begin{equation}
{\cal E} =  2 \sqrt{2\gamma G}\,E\,,
 \label{giving}
 \label{enescal}
 \end{equation}
re-expressing QES spectrum in terms of eigenvalues of asymmetric
two-diagonal matrix
 \begin{equation}
 \label{trapas}
 {H}(\mu)=
 \left(  \begin{array}{ccccc}
0 & f_1(\mu) & & &  \\
 f_{(N-1)}(0) & 0& f_2(\mu) &  &  \\
 &\ddots&\ddots&\ddots&\\
&& f_2(0) & 0&  f_{({N-1})}(\mu)   \\ &&& f_1(0) & 0
\end{array} \right)
 \end{equation}
where we abbreviated
 \be
 f_n(\mu)=n-(N-2n)\,\mu\,n\,, \ \ \ \ \ \ \ \ \ \ \
 \mu = \frac{1}{G}=\frac{1}{2m+N+D-2}\,.
 \label{measure}
 \ee
This matrix will now be understood as our QES ``Hamiltonian"
which is to be diagonalized perturbatively.  In a way guided by
a few experiments made at the first few lowest dimensions $N$,
we shall reveal and describe a new and efficient recipe for such
a construction in the next section.

\section{Perturbation expansions in $\mu^2$
 \label{treti}}

\subsection{Optimal shift: matrix dimension $N = 3$
  \label{prvnitreti}}

In our notation where the spatial dimension $D$ and all the
angular momenta $\ell=(D-3)/2+m$ with $m = 0, 1, \ldots$ are
``very" large, $D \gg 1$, the integer $N$ denotes just the {\em
matrix} dimension in eq. (\ref{tridSE}) and its numerical value
may be arbitrary.  Thus, our perturbation expansions will use
the smallness of $\mu = 1/G$ or, in a more general setting, of
$\mu'= 1/[D + 2m+ \delta(N)]$ where we have a full freedom in
the choice of an $N-$dependent shift $\delta=\delta(N)$.

Let us select $\delta(N)=1+\varepsilon$ at $N=3$, and reduce the
diagonalization of our ``Hamiltonian" (\ref{trapas}) to the
elementary secular equation
\[
\det\, \left [ \begin {array}{ccc}
E&1-\mu\,(1+\varepsilon)&0\\\noalign{\medskip}2&E&2+2\,\mu\,
(1-\varepsilon)
\\\noalign{\medskip}0&1&E\end {array}\right ]=0\,.
\]
Although its explicit polynomial form
${E}^{3}-4\,E+4\,E\,\mu\,\varepsilon=0$ is virtually trivial, it may
be still simplified and made perturbation independent, provided only
that we choose the above shift $\delta$ with the special and unique
$\varepsilon=0$. After this ``optimal" choice, all the three
zero-order energy roots $E_j\in \{-2,0, 2\}$ remain safely real and
become manifestly independent of the spatial dimension $D$. As a
consequence, all the higher-order corrections vanish at
$\varepsilon=0$, and the only $D-$dependence of the energies ${\cal
E}$ remains encoded in the above-mentioned scaling rule
(\ref{enescal}).

\subsection{Continued fractions:
matrix dimension $N=4$   \label{druhatreti}}

In the next case at $N=4$ one reveals that with another ``optimal"
$\delta(N)=2$,
an explicit linear $\mu-$dependence disappears again
from the secular equation
 \ben
  \det\, \left [ \begin {array}{cccc} {
 E}&1-2\,{\mu}&0&0
\\\noalign{\medskip}3&{E}&2&0\\\noalign{\medskip}0&2&{E}&3(1+2
 \,{\mu})\\\noalign{\medskip}0&0&1&{E}\end {array}\right ]
 =0\,
 \een
Only
the quadratic term persists
in the secular polynomial ${{ E}}^{4}-10\,{{ E}}^{2}+9-36\,{{
\mu}}^{2} $.
Thus, the choice of
 $N=4$ should be understood
as ``the first nontrivial" model. Not nontrivial enough of
course: the quadruplet of the secular roots keeps its closed form,
$E_j\in \{\pm \sqrt {5 \pm 2\,\sqrt {4+9\,{{ \mu}}^{2}}} \} $.
Circumventing the standard and tedious perturbation recipes, all
these four roots may be Taylor-expanded giving {\em directly} all the
four perturbation expansions in the powers of $\mu^2$,
 \be
 E_1=-E_4=
3+{\frac
{3}{4}}{{\mu}}^{2}-{\frac{33}{64}}{{\mu}}^{4}+{\frac{309}{512}}{{\mu}}^{6}
-{\frac{14133}{16384}}{{\mu}}^{8}+{\frac{179643}{131072}}{{\mu}}^{10}
+O\left ({{\mu}}^{12}\right )\,, \label{byloji6}
 \ee
 \be
 E_2=-E_3=
1-{\frac {9}{4}}{{\mu}}^{2}-{\frac {81}{64}}{{\mu}}^{4}-{\frac
{2187}{512}}{{\mu}}^{6}-{\frac {137781}{16384}}{{\mu}}^{8}-{ \frac
{3601989}{131072}}{{\mu}}^{10}+O\left ({{\mu}}^{12}\right )\,.
\label{bude4}
 \ee
Due to the analyticity of these expressions in the complexified
variable $\mu$, all of these four series have, obviously, the
same and non-vanishing circle of convergence with the radius
$\mu_{max} = 1/2$ determined by the nearest branch point in the
complex plane of $\mu$.  Once we recollect that for any $m \geq 0$
and $D>0$ we always have $1/\mu >\delta(4)=2$, we may conclude
that for all our four shifted-large$-\ell$ perturbation series
at $N=4$, the circle of convergence is amply
sufficient to cover the whole domain of physical interest.

The overall square-root form of the ``energies" $E$ might have
been taken into account in this context. At least, one may
expect a simplification of the formulae when the squares of the
energies $\Omega=E^2$ are taken into consideration.
For example, in place of
the latter two series
(\ref{bude4}),
the ``unified" power-series formula reads
 \be
  \Omega(\mu^2)
 = E_{2,3}^2=
 1-{\frac {9}{2}}{{{\mu}}}^{2}+{\frac {81}{32}}{{{\mu}}}^{4}-{\frac
{729}{256}}{{{\mu}}}^{6}+O\left ({{{\mu}}}^{8}\right )\,.
 \label{mop}
 \ee
Its use improves the rate of convergence because we
removed one of the branch points, arriving at the new value
$\mu_{max}^{[improved]} = 2/3$ for the radius of convergence.
In a broader context, it is therefore not surprizing that the
underlying idea of expanding a suitable function of $E$ (in our
case, a square) rather than this observable quantity itself did
find non-trivial applications elsewhere~\cite{Rafa}.

The transition from $E$ to its square $\Omega$ inspires also a
slightly counterintuitive re-arrangement of our secular
polynomial in an {incompletely} factorized form where the small
corrections are manifestly separated,
 \ben
 \left (
 \Omega-9
 \right )
 \left (
 \Omega-1
 \right )
 =36\,\mu^2\,.
 \een
This defines an explicit {\em analytic} continued-fraction
re-arrangement of the complete, ``perturbed" $\Omega(\mu)=\Omega(0)
+\mu^2\omega(\mu)$. For example, once we select
$\Omega(0)=E_{1,4}^2=9$, we arrive at the easily verified identity
 \be
\Omega(\mu^2)= 5 + 2\,\sqrt {4+9\,{{ \mu}}^{2}}=9+
 \frac{{36\,\mu^2}}{
 8+
 {\frac{36\,\mu^2}{8+\ldots}}}\,.
 \label{cf}
 \ee
The backward conversion of the analytic continued fraction on
the right-hand side to the plain perturbative power series in
$\mu^2$ is routine~\cite{Wall} and almost as trivial as its
above-mentioned Taylor-series derivation from the
left-hand-side square root.

\subsection{Convergence of iterations: matrix dimension
 $N=5$   \label{tretitreti}}

Our experience gained from the preceding, completely and easily
solvable examples applies to all of the higher $N$'s in a more or
less straightforward manner. Firstly, by extrapolation we conjecture
(and subsequently verify) that the ``optimally shifted"
 small
parameter $\mu$
 is {\em always} defined by eq. (\ref{measure}). Secondly, we pick up the
next $N=5$ and re-write the secular equation
 \ben
 \det\,\left [ \begin {array}{ccccc} {{E}}&1-3\,{\mu}&0&0&0
\\\noalign{\medskip}4&{{E}}&2(1-{\mu})&0&0\\\noalign{\medskip}0&
3&{{E}}&3(1+{\mu})&0\\\noalign{\medskip}0&0&2&{{E}}&4(1+3\,
\mu)\\\noalign{\medskip}0&0&0&1&{{E}}\end {array}\right ]=0
 \een
in the polynomial form
${{{E}}}^{5}-20\,{{{E}}}^{3}+64\,{E}-288\,{{E}}\,{{\mu }}^{2}=0$
showing that we may factor the zero root $E=E_3=0$ out. Later on we
shall see that such a root exists (and will be dropped) at all the
odd $N$. The fact that one of our roots is identically vanishing and
may be factored out, strengthens the parallels between $N=4$ and
$N=5$. In the latter case we get a quadruplet of the nontrivial
roots,
 \ben
 E_{1}=-E_5=
 \sqrt {10+ 6\,\sqrt {1+8\,{{\mu}}^{2}}}=
4+3\,{{\mu}}^{2}-{\frac {57}{8}}{{\mu}}^{4}+{\frac {939}{32}}{{
 \mu}}^{6}-{\frac {75957}{512}}{{\mu}}^{8}+O\left ({{\mu}}^{10 }
 \right )\,,
 \een
 \ben
 E_{2}=-
 E_{4}=
 \sqrt {10- 6\,\sqrt {1+8\,{{\mu}}^{2}}}=
 2-6\,{{\mu}}^{2}+3\,{{\mu}}^{4}-39\,{{\mu}}^{6}+{\frac {483}
 {4}}{{\mu}}^{8}-{\frac {3693}{4}}{{\mu}}^{10}+O\left ({{\mu}}^{12}
 \right )\,.
 \een
In full parallel with $N=4$ it is easy to show that $\mu_{max} =
1/(2\,\sqrt{2})$. This time, the domain of the perturbative
convergence covers more than we really need. On its boundary we set
$m=0$ and $D=1$ and get $\mu = \mu_{max\ phys} = 1/4 <
1/(2\,\sqrt{2}) =\mu_{max\ math}$.  In comparison with $N=4$ a
tendency is observed towards and improvement of the overall rate of
convergence, which are good news in the light of our intention to
move towards any $N$ in principle.

Let us once more recollect the existence of the $N=4$ continued
fractions for $\Omega$'s. It is related to the partial factorization
of the secular equation which may be easily transferred to $N=5$,
 \be
 \left (
 \Omega-16
 \right )
 \left (
 \Omega-4
 \right )
 =288\,\mu^2\,.
 \label{tohle}
 \ee
Re-scaling the measure of perturbation $\mu^2=\lambda/72$ and picking
up the zero-order root $\tilde{\Omega}^{[0]}=4$ for definiteness, the
corrections in $\Omega(\mu^2) =4\,( 1-Z ) $ may be defined by
iterations of eq. (\ref{tohle}), $Z_{[new]}=\lambda/(3+ Z_{[old]})$,
and they coincide again with an analytic continued fraction,
 \be
 Z = \frac{\lambda}{3+\frac{\lambda}{3+ \ldots}}.
 \label{cfb}
 \ee
The proof of its convergence
is easy when performed
 in the spirit of the
fixed-point method of ref. \cite{fixedpoint}.
Indeed,
the mapping
$Z_{[old]})\to Z_{[new]}$
has just two fixed points, the values of which are
known and given by the
quadratic equation
$Z_{[FP]}^{(\pm)}=Z_{[old]})= Z_{[new]}$.
In the next step one imagines that
in the vicinity of these two points,
the obvious sufficient condition for the
convergence/divergence of interations
of our mapping
(i.e., of our continued fraction with constant coefficients)
 is that the
derivative $Y(z)=Z_{[new]}'$ (of the dependent variable $Z=Z_{[new]}$
with respect to the independent variable $z=Z_{[old]})$) is
smaller/bigger than one in its absolute value, respectively
\cite{fixedpoint,fixedpointb}. We have all the necessary quantities
at our disposal so that the criterion is verified by trivial
insertions. In perturbative regime of the smallest $\lambda \ll 1$ it
is sufficient to specify the leading-order estimates for the small
positive $Z_{[FP]}^{(+)}=\lambda/12+\ldots$ and for the much larger
and negative $Z_{[FP]}^{(-)}=-3-\lambda/3+\ldots$. We may conclude
that the large fixed point is unstable since $Y^{(-)}= -9/\lambda +
\ldots \ll -1 $ while the small fixed point is stable since $Y^{(+)}=
-\lambda/9 + \ldots \in (-1,1) $.

We described this proof of convergence of continued fractions in full
detail because its idea is easily transferred to all the higher
dimensions $N$.  In the language of fixed points (or accumulation
points), one can also better understand the transition to the
power-series perturbation expansions of our squared energies
$\Omega(\mu^2)$. They may be re-constructed directly from the
recurrences \cite{fixedpointb}, without any recourse to the closed
square-root formulae.

\subsection{$N=6$ and the branching
 continued fractions   \label{ctvrtatreti} \label{hle6}}

At $N=6$ the extraction of the roots from the secular equation
 \be
  -225+259\,{{{E}}}^{2}
  -35\,{{{E}}}^{4}
 +{{{E}}}^{6}
  +3600\,{{{\mu}}}^{2}
  -1296
 \,{{{\mu}}}^{2}{{{E}}}^{2}
 =0
 \label{je6}
 \ee
ceases to be easy. Fortunately,  the $D \to \infty$ asymptotic
analysis of this equation reveals that there still exists an {\em
elementary} {partial} factorization of this equation,
 \be
 (E+5)\,(E+3)\,(E+1)\, (E-1)\,(E-3)\,(E-5)
 =144\,qm^2\,(3\,E-5)\,(3\,E+5).
 \label{bude6}
 \ee
Both the existence and the simplicity of this relation opens to us
new horizons. We realize that we may succeed in an algebraic
determination of the {\em perturbation representations of the roots}
in an almost as closed form as above. What is now vital is that the
construction {does not} still require any explicit knowledge of the
roots and might remain feasible at the higher $N$'s.

Once we return to eqs. (\ref{cf}) and (\ref{cfb}) for guidance, we
find that a specific generalization of the continued fractions exists
at $N=6$ since Firstly, perturbation anzatz $\Omega(\mu^2)
=1+\mu^2\,\omega (\mu^2)$ converts
 equation (\ref{bude6}) into iterative
recipe
 \be
\omega_{[new]}^{(k)}
= -
\frac{144\,(25 - 9\,\Omega)}
{(9-\Omega)(25-\Omega)}, \ \ \ \ \ \ \ \ \
\Omega= 1 + \mu^2\,
\omega_{[old]}^{(k)}\,, \ \ \ \ \ \ k = 0, 1, \ldots\,.
\label{vdova}
\ee
In comparison with the
continued-fraction predecessors
of this nonlinear two-term recurrence,
just a ``branching" of the continued fraction
occurs,
 \be
\omega_{[new]}^{(k)}
= 72\,\left (
\frac{7}{8-\mu^2\omega_{[old]}^{(k)}}
-\frac{25}{24-\mu^2\omega_{[old]}^{(k)}}
\right )
\,, \ \ \ \ \ \ k = 0, 1, \ldots\,.
\label{vdovaby}
\ee
These formulae generate the sequence of the values
$\omega_{[old]}^{(k+1)}=\omega_{[new]}^{(k)}$ with $k = 0, 1,
\ldots$ starting from the initial $\omega_{[old]}^{(0)}=0$
and giving
 \be
\omega_{[new]}^{(0)}=-12, \ \ \
\omega_{[new]}^{(1)}
= -
\frac{12\,(4+27\,\mu^2)}
{(2+3\,\mu^2)(2+\mu^2)}, \ \ \ldots\,.
\label{stepby}
 \ee
In the limit $k \to \infty$ (whenever it exists), one gets the
function $\omega_{[old]}^{(\infty)}=\omega_{[new]}^{(\infty)}$ of
$\mu^2$ whose form is a generalization of the standard analytic
continued fraction.  Its dependence on $\mu^2\geq 0$ is smooth. From
the negative value $\omega(0)=-12$ in the origin this function
decreases to a minimum $\omega(0.400186)=-24.94120$ and then it very
slowly grows again.

Whenever necessary, one can establish a number of parallels between
iterations (\ref{vdovaby}) and the current theory of continued
fractions \cite{Wall}. In particular, the fixed-point pattern of the
proofs of convergence of ref. \cite{fixedpoint} applies to functions
$\Omega(\mu^2)$ at both $N=4$ or $5$ and $N=6$ (and, as we shall see
below, $7$). Thus, for the ``first nontrivial" $N=6$ mapping
$\omega_{[old]}^{(k)} \longrightarrow \omega_{[new]}^{(k)}$, we may
skip the details (like, e.g., all the small$-\lambda$ analysis) and
emphasize only that the proof of the convergence of our branched
continued fraction consists of three steps now. Firstly, one
demonstrates that there is just one fixed-point root which is
compatible with the perturbation smallness of corrections. Secondly,
one confirms that this ``only acceptable" fixed point is stable (by
showing that the derivative of the mapping at this point is
sufficiently small, $|Y(z)| < 1$). Finally, one shows that the stable
point is unique, which follows from the observation that $|Y(z)|>1$
at the other two fixed points.

For giving to the reader a rough estimate of the rate of convergence
of the iterations, let us pick up a sample value of the parameter
$\lambda=\mu^2=1/10$, giving the fairly small $Y(z)\approx 0.25949$
at the accumulation point $z \approx -2.30666$, while the quick
divergence of the iterations at the other two fixed points results
from the large magnitude of the derivatives  $Y(5.26698)\approx
3.0556$ and $Y(37.03969) \approx -5.5675$ there.

For many practical purposes, it is desirable to convert our branched
continued fractions
$\omega_{[old]}^{(\infty)}=\omega_{[new]}^{(\infty)}$ in example
(\ref{vdova}) into the more common Rayleigh-Schr\"{o}dinger-type
power series.  We performed such a conversion giving, for the squared
energy, the following perturbation series in $\mu^2$,
 \ben
\Omega=1-{12}\mu^2-{57}\mu^4-{\frac {591}{4}}\mu^{6}+{\frac {4215}
{16}}\mu^{8}+{\frac {286293}{64}}\mu^{10} +
\een
 \be
+{\frac {3702951}{256}}\mu^{12}-{
\frac {63786951}{1024}}\mu^{14}-{\frac {3242255193}{4096}}\mu^{16}-
\label{23}
\ee
 \ben
-{\frac { 32707656915}{16384}}\mu^{18}+{\frac
{1182033909831}{65536}}\mu^{20}+ {\cal O} \left (\mu^{22}\right ) \,.
 \een
Up to the order ${\cal O} \left (\mu^{26}\right )$ we verified the
validity of an empirical rule that the signs are changing after every
third order. {\it A posteriori}, one may expect an efficient
numerical summability of this series by the standard
Pad\'{e}-resummation techniques. We skip here the really amusing
possibility of a backward comparison of the resulting rational
approximations with their available {\em initial}
versions~(\ref{stepby}).

\subsection{Expansions at an arbitrary matrix dimension $N$
   \label{patatreti}}

In a way which is outlined in Table~1 and which may be tested on
the further non-numerical QES constructions (with dimensions $N=7$ --
$N=9$, i.e., up to quartic-polynomials) as well as at any higher
$N$, we have reached the stage where the overall structure of
secular polynomials is clear,
\be
\ {\cal P}(\Omega)
=
{\cal P}_0(\Omega)
4sy+\mu^2
{\cal P}_1(\Omega)
+ \ldots + \mu^{2K}
{\cal P}_K(\Omega)\ , \ \ \ \ \ \ \
K = entier
\left [
\frac{N}{4}
\right ]\,.
\ee
The routine analysis confirmed our expectations that at
any $N$, the large parameters $\ell$ or $D$ entering the measure
of smallness $\mu$ as defined by eq. (\ref{measure}) are
``optimally shifted". This assertion is supported by the
following four reasons at least.

\begin{itemize}

\item
All the odd powers of $\mu$ disappear from the
polynomial forms of the secular determinants as well as from the
perturbation expansions of the observable quantities (energies).

\item
Up to the dimension $N=9$ which is already fairly large, the
degree of our ``optimalized" secular polynomials does not exceed
four; this means that their factorization may be made
non-numerically.

\item
At all the larger $N>9$, a partial (i.e., asymptotic, $D \to
\infty$) factorization is still feasible
non-numerically in zero order.
As a consequence, all the higher-order ${\cal O}(\mu^2)$
corrections may be generated in the form which, in many a
respect, generalizes the analytic
continued fraction.

\end{itemize}

\noindent Surprizingly enough, the use of our present generalizations
of continued fractions has led to a very transparent alternative to
Cardano formulae at dimension as low as $N=6$. Similarly, at $N=7$ we
eliminate the trivial root $E=0$ and get the rule
  \ben
 -2304\,{{E}}
 +
 784\,{{{E}}}^{3}
 -56 \,{{{E}}}^{5}
 +{{{E}}}^{7}
 +
 40320\,{{{\mu}}}^{2}{{E}}
 -4320\,{{{\mu}}}^{2}{{{E}}}^{3}
 =0
 \een
and the parallel update of eq. (\ref{bude6}),
 \be
  \left (
\Omega - 2^2 \right )
   \left (
\Omega - 4^2 \right )
    \left (
\Omega - 6^2
 \right )
 =1440\,\mu^2\,(-28+3\,\Omega)\,.
\label{bude7}
 \ee
We need not repeat the extraction of the conclusions similar to the
ones in section~\ref{hle6}. A not too dissimilar remark may be added
concerning $N=8$ with
 \be
  \left (
\Omega - 1^2 \right )
  \left (
\Omega - 3^2 \right )
   \left (
\Omega - 5^2 \right )
    \left (
\Omega - 7^2
 \right )
 =
360\,\mu^2\, \left ( 1225- 682\,\Omega + 33\,\Omega^2 \right ) +
1587600\,\mu^4\,. \label{bude8}
 \ee
This equation illustrates both the emergence of the higher powers of
$\mu^2$ on the right-hand side {\em and} the parallels with the $N=9$
secular equation
 \ben
  \left (
\Omega - 2^2 \right )
  \left (
\Omega - 4^2 \right )
   \left (
\Omega - 6^2 \right )
    \left (
\Omega - 8^2
 \right )
 =
 \een
 \ben
 = 288\,\mu^2\, \left ( 22016- 3740\,\Omega + 99\,\Omega^2 \right ) +
24385536\,\mu^4\, \label{bude9}
 \een
as well as an unexpectedly smooth character of transition  to the
``first unsolvable" $N=10$ example
 \ben
-893025
+1057221\,{{{E}}}^{2}
-172810\, {{{E}}}^{4}
+8778\, {{{E}}}^{6}
-165\,{{{E}}}^{8}
+{{{E}}}^{10}
+ \een  \ben
+71442000\,{{\mu}}^{2}
-48647664\, {{\mu}}^{2}{{{E}}}^{2}
+3809520\,{{\mu}}^{2} {{{E}}}^{4}
-61776\,{{\mu}} ^{2} {{{E}}}^{6}
- \een  \ben
-914457600\,{{
\mu}}^{4}
+199148544\,{{\mu}}^{4}{{{E}}}^{2}
\,.
 \een
does not bring, in full accord with the scheme of Table~1, anything
new, indeed.  The general pattern indicated in Table~1 works and
remains valid for all the matrix dimensions $N=4K, 4K+1, 4K+3$ and
$4K+3$ with any auxiliary $K=1,2,\ldots$.

On this basis, we are permitted to pay attention to any root
$\tilde{\Omega}$ of our zero-order secular polynomial ${\cal
P}_0(\Omega)=(\tilde{\Omega}-\Omega)\,{\cal Q}(\Omega)$.  As
long as this root is an integer, it is extremely easy to study
its small vicinity by the insertion of the ansatz
$\Omega=\tilde{\Omega}+\mu^2\omega(\mu^2)$ into the exact
secular equation $\ {\cal P}(\Omega)=0$.  The resulting new form
of our secular equation reads
\be
\mu^2\omega(\mu^2)= \frac{1}{ {\cal Q}_0(\Omega)}\, \left [ \mu^2
{\cal P}_1(\Omega) + \ldots + \mu^{2K} {\cal P}_K(\Omega) \right ]\,.
\label{perd} \ee On the basis of this formula we may consider a
sequence $\omega_j(\mu^2) $ of approximate corrections initiated by
$\omega_{-1}(\mu^2) =\omega_{-2}(\mu^2) = \ldots = 0$. The use of the
abbreviations $\Omega_j=\tilde{\Omega}+\mu^2\omega_j(\mu^2)$
re-interprets finally our secular equation (\ref{perd}) as an
innovated and most important recurrent relation
 \be
\omega_{j}(\mu^2)= \frac{{\cal P}_1(\Omega_{j-1}) }{ {\cal
Q}_0(\Omega_{j-1})} + \ldots + \mu^{2K-4}\,\frac{{\cal
P}_{K-1}(\Omega_{j-K+1}) }{ {\cal Q}_0(\Omega_{j-K+1})} +
\mu^{2K-2}\,\frac{{\cal P}_K(\Omega_{j-K}) }{ {\cal
Q}_0(\Omega_{j-K})} \,, \ \ \ \ \ \ j = 0,1, \ldots\,.
\label{perdoch}
 \ee
At $K=1$ and $N=4$ or $N=5$ this recipe returns us back to the
current definition of the analytic continued fractions. Similarly, at
$N=6$ and $N=7$ we get their branched alterantive. At all the higher
$N \geq 8$, our recurrences (\ref{perdoch}) offer just an immediate
rational generalization of the latter two concepts.

\section{Interpretation and outlook \label{ctvrta}}

\subsection{Zero-order Schr\"{o}dinger equation\label{prvnictvrta}}

We have seen that once we switch to the rescaled energy variable $E$
in eq. (\ref{enescal}) and to the Hamiltonian (\ref{trapas}), the
zero-order diagonalization of $H(0)$ becomes unexpectedly easy at
{\em any} $N$. In spite of the manifest asymmetry of the
Schr\"{o}dinger QES equation in the limit $\ell \to \infty$,
 \begin{equation}
 \label{trap}
 \left(  \begin{array}{ccccc}
0 & 1 & & &  \\
 (N-1) & 0& 2 &  &  \\
 &\ddots&\ddots&\ddots&\\
&& 2 & 0&  ({N-1})   \\ &&& 1 & 0
\end{array} \right)
 \left(  \begin{array}{c}
 {p}_0\\
 {p}_1\\
\vdots \\
 {p}_{N-2}\\
 {p}_{N-1}
\end{array} \right)
=
E \cdot
 \left(  \begin{array}{c}
 {p}_0\\
 {p}_1\\
\vdots \\
 {p}_{N-2}\\
 {p}_{N-1}
\end{array} \right )\,,
 \end{equation}
all its eigenvalues remain strictly real, equal to integers
and equidistant,
  \begin{equation}
\left ( E_1, E_2, E_3, \ldots, E_{N-1}, E_N \right ) = \left (
-N+1, -N+3, -N+5,  \ldots, N-3, N-1 \right ).
 \label{mainr}
 \end{equation}
It is quite elementary to verify that also the the respective left
and right eigenvectors of $H(0)$ remain real. Up to their norm, all
of them can be represented in terms of integers as well. Their
components may be arranged in the rows and columns of the following
$N$ by $N$ square matrices $P=P(N)$,
 \[
 P(1) = 1, \ \ \ \ \ \ \ P(2)=
\frac{1}{\sqrt{2}} \,
 \left(  \begin{array}{rr}
1&1\\ 1&-1
\end{array} \right),
 \]  \[  P(3)= \frac{1}{\sqrt{4}} \,
 \left(  \begin{array}{rrr}
1&1&1\\ 2&0&-2\\ 1&-1&1
\end{array} \right),
 \ \ \ \ \ \ \
P(4)= \frac{1}{\sqrt{8}} \,
 \left(  \begin{array}{rrrr}
1&1&1&1\\ 3&1&-1&-3\\ 3&-1&-1&3\\ 1&-1&1&-1
\end{array} \right),
 \]
 \[  P(5)= \frac{1}{\sqrt{16}} \,
 \left(  \begin{array}{rrrrr}
1&1&1&1&1\\ 4&2&0&-2&-4\\ 6&0&-2&0&6\\ 4&-2&0&2&-4\\ 1&-1&1&-1&1
\end{array} \right), \ldots\,.
 \]
All these matrices are asymmetric {\em but} idempotent, $P^2=I$. This
implies that the Hamiltonian in our zero-order QES sextic
Schr\"{o}dinger equation $H^{(0)} \vec{p}^{(0)} =\vec{p}^{(0)}
{E}^{(0)} $ need not be diagonalized at all. Indeed, as long as the
$N-$plets of the zero-order (lower-case) vectors $ \vec{p}^{(0)} $
are concatenated into the above-mentioned (upper-case) $N$ by $N$
matrices $P=P^{(0)}$, we may also collect all the pertaining
eigenvalues ${E}^{(0)}$ in a diagonal matrix ${\varepsilon}^{(0)}$.
In this arrangement, the unperturbed Hamiltonian is factorized,
$H^{(0)} = P{\varepsilon}^{(0)}P$. As a consequence, the zero-order
equation is an identity since, in our compactified notation, it reads
$ P{\varepsilon}^{(0)}PP = P{\varepsilon}^{(0)}$ and we know that
$P^2=I$. In the next section we show that and how similar notation
may be used in all orders.

\subsection{Rayleigh-Schr\"{o}dinger perturbation
recipe revisited \label{druhactvrta}}

We have seen that

At any finite value of the spatial dimension $D$ we have seen in
section \ref{treti} that the routine power-series ansatz of
perturbation theory becomes applicable even though the unperturbed
Hamiltonian $H(\mu)$ itself is non-diagonal. We may write
 \[
H(\lambda)=H^{(0)} + \lambda\,H^{(1)}, \ \ \ \
\ \lambda = \mu^2
 \]
where the perturbation is an asymmetric one-diagonal matrix in our
particular illustrative example. Even without the latter constraint
we arrive at the textbook perturbative representation of our matrix
Schr\"{o}dinger equation,
 \begin{equation}
 \begin{array}{c}
\left ( H^{(0)} + \lambda\,H^{(1)}  \right )
\cdot \left ( {p}^{(0)} + \lambda\,{p}^{(1)} + \ldots +
 \lambda^K {p}^{(K)}
+{\cal O}(\lambda^{K+1}) \right )\\
=
\left ( {p}^{(0)} + \ldots +
 \lambda^K {p}^{(K)}
+{\cal O}(\lambda^{K+1}) \right ) \cdot \left ( {E}^{(0)}
 +
\ldots +
 \lambda^K {E}^{(K)}
+{\cal O}(\lambda^{K+1}) \right ). \end{array} \label{pertu}
 \end{equation}
The sets of the vectors for corrections $\vec{{p}}^{(k)}_j, \,j=1, 2,
\ldots, N$ may be concatenated in the square matrices $\Psi^{(k)}$.
This enables us to re-write the first-order ${\cal O}(\lambda)$ part
of eq. (\ref{pertu}) in the particularly compact matrix form
 \begin{equation}
{\varepsilon}^{(1)} + P\,\Psi^{(1)} {\varepsilon}^{(0)}
-
{\varepsilon}^{(0)} P\, \Psi^{(1)}
=
P\,H^{(1)}\,P. \label{prvnirad}
 \end{equation}
In the second order we get
 \begin{equation}
{\varepsilon}^{(2)} + P\,\Psi^{(2)} {\varepsilon}^{(0)}
-
{\varepsilon}^{(0)} P\, \Psi^{(2)}
=
P\,H^{(2)}\,P + P\,H^{(1)}\,\Psi^{(1)}
-
P\,\Psi^{(1)} {\varepsilon}^{(1)} \label{druhyrad}
 \end{equation}
etc.  This is a hierarchy of equations representing their source
(\ref{pertu}) order-by-order in $\lambda$. Their new merit lies in
their recurrent character. Their ``old" or ``input" data occur on the
right-hand side of these equations, while the ``new" or ``unknown"
quantities stand to the left. All the higher-order prescriptions have
the same structure. In all of them, the diagonal part of each
equation (i.e., of (\ref{prvnirad}) or (\ref{druhyrad}) etc)
determines the diagonal matrices containing energy corrections (i.e.,
${\varepsilon}^{(1)}$ or ${\varepsilon}^{(2)}$ etc, respectively).
Non-diagonal components of these matrix relations are to be
understood as definitions of the eigenvectors, with an appropriate
account of the well known normalization freedom which has been
thoroughly discussed elsewhere \cite{Czechdva}. All these relations
just re-tell the story of our preceding section but after an
appropriate modification they may also be used for the evaluation of
separate corrections in some more complicated QES
systems~\cite{quartic}.

In the conclusion, we may emphasize that the practical reliability of
any perturbation prescription is, mostly, determined by the quality
of the zero-order approximation. In this sense, our present study
offers also a broadening of their menu. In the light of our results
one will be forced to make a more careful selection between the ES
and QES $V^{[(Q)ES]}(|\vec{r}|)$. Indeed, in each of these respective
extremes one encounters different difficulties. Still, the {\em
simpler} ES-based choice of $V_0(|\vec{r}|)$ is {\em predominantly}
preferred in practice. Almost without exceptions, such a decision is
being made for one of the following three apparently good reasons.

\begin{itemize}

\item
Up to now, many QES-based constructions stayed within the mere
lowest-order  perturbation  regime since the majority of the naive
implementations of the Rayleigh-Schr\"{o}dinger perturbation scheme
becomes complicated in the high orders \cite{Czech}.

\item
{\it A priori}, any  available set of the QES bound states $\psi_0$
is, by definition, incomplete. This makes the matrix form of the
pertaining propagators non-diagonal and, hence, not too easy to use
even in the lowest orders.

\item
Last but not least, many ambitious QES models cannot be used as
eligible for perturbations because they are almost frighteningly
complicated even in zero order~\cite{Ushveridze,Leach}.

\end{itemize}

 \noindent
Within our new perturbation prescription, all these shortcomings were
at least partially weakened. At the same time, the appeal of the
ES-based perturbation constructions should at least slightly be
re-evaluated as well: (i) the class of the available
$V^{[ES]}(|\vec{r}|)$ is really extremely narrow for many practical
purposes; (ii) several important (e.g., double-well) models lead to a
perturbation which is not ``sufficiently small" in the sense of Kato
\cite{Kato}; (iii) one needs to go to the really very high
perturbations in many cases of practical interest~\cite{Cizek}.

In such a setting we showed that the {\em technical} treatment of the
QES-type zero-order approximations {\em can} be significantly
simplified. This could help, say, in situations where we have to deal
with a phenomenological potential $V(r)$ which cannot be well
approximated by {any available} exactly solvable
$V^{[ES]}(|\vec{r}|)$ while there exists an exceedingly good
approximation of $V(r)$ by a partially solvable
$V^{[QES]}(|\vec{r}|)$. Under this assumption we may recommend the
present construction as a guide to new applications of the old
large$-\ell$ expansion idea~\cite{jednadel}, especially when a very
high order $N$ characterizes the polynomial part of the QES wave
function $\psi^{[N]}(r)$, which would makes the simple-minded
zero-order construction prohibitively complicated by itself.

\section{Summary \label{sesta}}

The problem of construction of large QES multiplets has been
addressed here within the framework of the so called large$-\ell$
expansion method. Its key idea is very popular and re-emerges
whenever radial Schr\"{o}dinger equation is considered at a large
angular momentum $\ell$.  Originally, this attracted our attention
because in all these recipes  there exists an obvious {\em ambiguity}
in  a {\em free} choice between alternative small parameters
$\lambda=1/\ell$ and $\lambda_{shifted}=1/(\ell+\beta)$. We should
remind the reader that there exist in fact many alternative
shifted-large$-\ell$ versions of the expansions which use this
freedom in different ways \cite{shiftedel} but, mostly, this
parameter plays a certain not too essential variational role. In this
context, the position of our present approach may be exceptional; in
QES context the role of our ``optimal" and unique $\beta$ happened to
be much more essential, influencing not only the rate of convergence
but also, directly, the form of the perturbation series itself.

Thus, although in most cases the transition to a shifted large$-\ell$
expansion is being designed to extend its practical applicability, we
imagined that the construction of the $N-$plets of quasi-exact
sextic-oscillator bound states $\psi_n^{[N]}(r)$ {\em does not admit}
any free choice of the shift $\beta$ at all.  On the contrary, our
approach has been shown to prefer the {\em unique}, ``optimal"
definition of the shift $\beta=\beta(N)$ which  leads to one of the
best available constructions of the QES spectrum. This is our main
result.

Our secondary motivation stemmed from the fact that whenever one
moves beyond the first few trivial multiplet dimensions $N$, the
practical appeal of QES solutions fades away.  In this sense we have
verified once more that one of the most efficient strategies of
solving ``difficult" Schr\"{o}dinger equation is provided by the
standard Rayleigh-Schr\"{o}dinger perturbation series~\cite{Kato} and
by many of its various practical modifications and alternative
implementations~\cite{Constantinescu}.  We succeeded in developing a
prescription with several merits described throughout the text.

In a broader methodical setting we were guided by the idea that the
quality of perturbative results depends on the two decisive factors,
viz., on the number of the available perturbation corrections and on
the rate of convergence of their series.  In both these aspects our
new approach passes the test very well. We have seen that in
comparison with the standard approach summarized for our present
purposes in section \ref{druhadruha}, we arrived at the more compact
and, presumably, most easily generated representation of our QES
energies. Moreover, in a sharp contrast to the manifest divergence of
the former (though, admittedly, more universal) recipe
\cite{Bjerrum}, our algebraic approach represents the QES spectra by
{\em convergent} infinite series.

Our new implementation of some old ideas of perturbation theory looks
promising and will definitely serve as a guide to the more
complicated QES constructions in the future.  For the sake of
simplicity we have chosen just one of the simplest models where the
sextic oscillator is not even complemented by a quartic force and
admits merely a quadratic repulsion near the origin.  This simplified
many technicalities {\em and} strengthened our belief that any future
transition to models with more parameters may and will proceed along
the same lines.

\subsection*{Acknowledgements}

Work supported by the GA AS CR grant Nr. A 104 8302.

%\newpage

\newpage

Table 1.
QES secular polynomials $\ {\cal P}(\Omega)
=
{\cal P}_0(\Omega)
+\mu^2
{\cal P}_1(\Omega)
+ \ldots + \mu^{2K}
{\cal P}_K(\Omega)\ $
\hspace{1.5cm} (the trivial factor $E$ ignored at odd $N$).

$$
\begin{array}{||c||c||c|c|c|c||}
\hline
\hline
{\rm dimension}
&{\rm perturbation}
&
\multicolumn{4}{c||}{{\rm
degree\ of\ } {\cal P}_j(\Omega)
}\\
\hline
 N&{\rm in\ } {\cal P}(\Omega)&
j=0&j=1&j=2&j=3\\
\hline
\hline
0\ {\rm and}\ 1
&{\rm absent}\ (K=0)& 0&-&-&-\\
2\ {\rm and}\ 3
&{\rm absent}\ (K=0)& 1&-&-&-\\
4\ {\rm and}\ 5
&{\rm linear}\ (K=1)& 2&0&-&-\\
6\ {\rm and}\ 7
&{\rm linear}\ (K=1)& 3&1&-&-\\
8\ {\rm and}\ 9
&{\rm quadratic}\ (K=2)& 4&2&0&-\\
10\ {\rm and}\ 11
&{\rm quadratic}\ (K=2)& 5&3&1&-\\
12\ {\rm and}\ 13
&{\rm cubic}\ (K=3)& 6&4&2&0\\
\ldots &\ldots &&&&\\
\hline
\hline
\end{array}
$$

\end{document}